\documentclass[aps,pre,twocolumn,superscriptaddress,floatfix,nofootinbib]{revtex4-2}

\usepackage{graphicx}
\usepackage{amsmath,amssymb}
\usepackage{bm}
\usepackage[colorlinks=true,
            linkcolor=blue,
            citecolor=blue,
            urlcolor=blue]{hyperref}

\usepackage{microtype}
\begin{document}

\title{Effect of Noise on Spatio-Temporal Evolution of Current Filamentation Instability in Relativistic Beam-Plasma Systems}

\author{Thulasidharan K}
\affiliation{School of Advanced Sciences, Vellore Institute of Technology, Vellore, India}

\author{Vishwa Bandhu Pathak}
\email[Corresponding author: ]{vishwa.bandhu@vit.ac.in}
\affiliation{School of Advanced Sciences, Vellore Institute of Technology, Vellore, India}

\begin{abstract}
The spatio-temporal evolution of the current filamentation instability in a
relativistic beam--plasma system is studied analytically and with two-dimensional particle-in-cell simulations. A partial differential equation for the transverse vector potential is derived for a sharp-front relativistic beam entering cold, unmagnetized plasma, including a second-order spatial derivative term that governs the spatial growth near the beam front. The equation is solved analytically for constant, linearly growing, and oscillatory initial noise when this term is neglected, and numerically when it is included, as no closed-form solution then exists. For constant initial noise, the numerical solution reproduces the simulated magnetic-field structure, unlike the analytical solution without the term. This shows that the longitudinal field modulation is intrinsic to the instability, present even for a noise with constant amplitudes. The noise profile as well can influence the spatial-temporal evolution of the instability, which we discuss further considering linearly growing and oscillatory noise. The field grows spatially behind the beam front and saturates at a length $L_{\mathrm{sat}}\propto(v_{0b}+2)v_{0b}/\gamma_{0b}^{3}$,
where $v_{0b}$ and $\gamma_{0b}$ are the beam velocity and Lorentz factor, beyond which growth is purely temporal. The saturation length increases linearly in time at a constant rate $\mathrm{d}L_{\mathrm{sat}}/\mathrm{d}\tau\approx0.42\,c$, matching the analytical estimate. The temporal growth rate remains unchanged, so the term modifies the spatial transport of the instability rather than its local amplification. For beam velocities above $0.6c$, the model deviates from the simulations as oblique modes and nonlinear filament dynamics outside the
single-mode treatment become important.
\end{abstract}

\maketitle

\section{Introduction}
When a relativistic charged-particle beam propagates through an unmagnetized background plasma, it introduces a net current in the background plasma and momentum anisotropy which acts as a source of free energy for electromagnetic and electrostatic perturbations. This leads to a broad spectrum of collective instabilities. These processes are encountered in laboratory experiments involving high-current relativistic electron beams as well as in astrophysical environments where energetic particle streams propagate through weakly magnetized or unmagnetized plasma, such as the interstellar medium surrounding gamma-ray burst sources and active galactic nucleus jets \cite{bret2010multidimensional,melrose1986,treumann2009}. Early theoretical work established that velocity-space anisotropy alone is sufficient to destabilize a plasma and drive spontaneously growing transverse perturbations even without an externally applied magnetic field \cite{rosenbluth1957,weibel1959,fried1959}. Later studies showed that beam--plasma systems simultaneously support several unstable branches, comprising longitudinal electrostatic two-stream modes together with the transverse electromagnetic Weibel and current filamentation modes, whose relative growth rates are set by the beam-to-plasma density ratio, the relativistic Lorentz factor, and the thermal spread of the distributions \cite{bret2009,bret2010,bret2004}.

Among the transverse electromagnetic instabilities in unmagnetized beam--plasma systems, the Weibel instability and the current filamentation instability (CFI) are primarily responsible for spontaneous magnetic-field generation \cite{weibel1959,fried1959}. Although both mechanisms produce transverse magnetic structures, they draw on different components of the available free energy. The Weibel instability is driven by anisotropy in the particle momentum distribution: unequal thermal or drift spreads in two orthogonal directions generate transverse current perturbations whose self-consistent magnetic field is amplified exponentially at the expense of the kinetic anisotropy \cite{yoon1987,davidson1972,califano1998}. The current filamentation instability, by contrast, originates from the transverse separation of the beam and return currents under their mutual Lorentz force: magnetic self-pinching reorganises the two current populations into isolated filamentary channels accompanied by localised magnetic-field growth \cite{bret2010,silva2003,yang1994}. In relativistic beam propagation these two processes frequently develop simultaneously, and their nonlinear interaction governs transverse current redistribution, filament coalescence, and the resulting broadband magnetic turbulence \cite{kazimura1998,stockem2014}. Particle-in-cell simulations have shown that this turbulence plays a central role in collisionless shock formation and non-thermal particle acceleration in relativistic astrophysical environments, including gamma-ray bursts and AGN jets \cite{medvedev2005,nishikawa2009,sironi2011,fiuza2020electron,frederiksen2004,hededal2004}.

The overwhelming majority of analytical treatments of CFI are formulated within a purely temporal framework. Perturbations are taken to vary as $\exp(i\vec{k}\cdot\vec{r}-i\omega t)$ in an infinite, spatially homogeneous plasma, the linearized dispersion relation is solved for the complex frequency $\omega(\vec{k})$, and the imaginary part yields a growth rate that is assumed identical at every point of the interaction volume \cite{yoon1987,bret2010,bret2009}. This approach is highly successful at predicting dispersion relations, maximum growth rates, and the dominant unstable wave numbers, and it underpins much of the present understanding of which mode prevails in a given parameter regime \cite{bret2009,bret2004}. It is, however, structurally incapable of describing how the instability is established in space. A physical beam possesses a leading edge that continuously encounters fresh, unperturbed plasma, so that the perturbation is seeded at the beam head and convected into the beam body rather than switched on uniformly throughout the volume. PIC studies of interpenetrating shells and relativistic shock precursors confirm this picture directly. In these simulations the generated magnetic field is not spatially uniform but remains concentrated behind the beam front, while the saturation boundary that separates the growing region from the saturated region migrates downstream as the system evolves \cite{silva2003,frederiksen2004,nishikawa2009,chang2008}. The instability is therefore convective near the beam head, and the spatial amplification phase precedes and conditions the subsequent temporal regime. This intrinsic coupling between propagation distance and instability growth, together with the finite extent of the magnetized region that it sets, cannot be recovered from a temporal dispersion relation, and it motivates an explicitly spatio-temporal description of CFI.

A reduced spatio-temporal model that captures this behaviour was introduced by Pathak \textit{et al.}~\cite{pathak2015spatial}. Working in the co-moving frame $\psi=v_{0b}t-z$, $\tau=t$, they reduced the wave equation for the transverse vector potential of a relativistic beam entering uniform cold plasma to a single partial differential equation for the amplitude $A(\psi,\tau)$. Their analysis showed that the field first develops as a spatially growing mode near the beam head and subsequently crosses over to a temporally dominated regime beyond a characteristic saturation length $L_{\mathrm{sat}}=Q\tau$, where $Q$ is a spatial-propagation coefficient set by the beam plasma frequency, the velocity, and the Lorentz factor. The model reproduced the early-stage spatial growth observed in PIC simulations and thereby provided the first compact description of CFI front propagation. The reduction was nonetheless carried only to leading order in the spatial coupling, so that the next term in the expansion, proportional to $\partial_\psi^2 A$ with coefficient $P=Qv_{0b}/2$, was discarded as small. This truncation constitutes the principal limitation of the leading-order model. For beam velocities in the intermediate-to-mildly relativistic range $0.4c$--$0.7c$, the ratio $P/Q=v_{0b}/2$ lies between $0.2$ and $0.35$, so that the neglected term represents a $20$--$35\%$ correction rather than a negligible one. The leading-order formulation consequently underestimates the spatial extent of the magnetized region, advances the saturation front too slowly, and omits the additional downstream modulation of the field profile that the simulations display.

The reason this omission cannot be left unexamined lies in the quantity that it controls. The physically decisive output of any spatio-temporal description of CFI is not the local growth rate but the length of the magnetized region behind the beam front, since it is this scale that enters estimates of the diffusive-acceleration mean free path, the width of the Weibel-mediated shock ramp, and the threshold for filament coalescence and the inverse magnetic-energy cascade \cite{achterberg2007,spitkovsky2008a,chang2008,silva2003,stockem2014}. In the leading-order model this scale is fixed by $Q$ alone, whereas the neglected term enters through the same coefficient $P=Qv_{0b}/2$, which reaches approximately a third of $Q$ for $v_{0b}\sim0.6c$. These are precisely the intermediate-to-mildly relativistic velocities that characterize the deceleration phase of gamma-ray burst ejecta and the downstream flow of mildly relativistic AGN jets, in which CFI is invoked to seed collisionless shock formation and non-thermal particle acceleration \cite{gehrels2002brightest,medvedev1999,spitkovsky2008particle,fiuza2020electron}. A precursor scale that is uncertain at the level of several tens of percent propagates directly into every quantity built upon it, and establishing the correct propagation speed of the saturation front is therefore a question of physical fidelity rather than of mathematical completeness.

In the present work the $P\,\partial_\psi^2 A$ term is restored, and its effect on the spatio-temporal evolution of the instability is examined. Analytical solutions of the reduced model are obtained for constant, linearly growing, and oscillatory inflow noise, of which the linear and oscillatory cases are derived here for the first time, and the extended model is integrated numerically for the same three profiles. Both formulations are then benchmarked against two-dimensional OSIRIS PIC simulations~\cite{fonseca2002} of a sharp-front relativistic beam entering cold plasma. The correction advances the saturation front more rapidly than the leading-order prediction and yields markedly closer agreement with the simulations for $v_{0b}\le0.6c$, while the temporal growth rate remains unaltered. Above this velocity the agreement deteriorates as oblique modes and nonlinear filament dynamics, which lie beyond the single-mode reduced description, become significant \cite{bret2004,spitkovsky2008a}. The remainder of the paper presents the derivation of the extended equation, its analytical and numerical solutions, and the quantitative comparison with simulation that establishes the modified saturation length together with its dependence on beam velocity and inflow noise.

\section{Theoretical Model}
\label{sec:const}

Consider a relativistic beam with velocity $v_{0b}$ and density $n_{0b}$ propagating in the $+z$ direction through a stationary plasma in which the ions are immobile and the background electron density is $n_{0p}$. Working in the co-moving frame defined by $\psi = v_{0b}t - z$ and $\tau = t$, the vector potential of the self-generated electromagnetic wave is written as $\vec{A} = A(\psi, \tau)\,e^{-ikx}\hat{z}$, which satisfies the Lorenz gauge condition. The associated magnetic and electric fields are $\vec{B} = \vec{\nabla}\times\vec{A}$ and $\vec{E} = -c^{-1}\partial_t\vec{A}$, respectively, and the vector potential obeys the wave equation $\left( \nabla^2 - c^{-2}\partial_t^2 \right)\vec{A} = -4\pi c^{-1} \vec{J}$. The current density is $\vec{J} = -e[n_b \vec{v}_b + n_p \vec{v}_p]$, where $n_b$ and $n_p$ are the beam and plasma number densities and $\vec{v}_b$ and $\vec{v}_p$ the corresponding velocities. As the beam enters the plasma, it perturbs both the density and the velocity of the beam and plasma electrons. The densities are therefore written as \(n_b=n_{0b}F(\psi)+n_{1b}\) and \(n_p=n_{0p}+n_{1p}\), where \(F(\psi)\) is the equilibrium beam density profile, and the velocities as \(\vec{v}_b=\vec{v}_{0b}+\vec{v}_{1b}\) and \(\vec{v}_p=\vec{v}_{1p}\). Retaining only the first-order perturbed quantities, the current density reduces to \(\vec{J}=-e[n_{0b}F(\psi)\vec{v}_{1b}+n_{0p}\vec{v}_{1p}+n_{1b}\vec{v}_{0b}]\). Substituting this expression, together with the assumed form of the vector potential, into the wave equation yields the evolution equation for the amplitude \(A(\psi,\tau)\),

\begin{equation}
\left[
\frac{1}{c^2}\partial_{\tau}^2
+
\frac{2v_{0b}}{c^2}\partial_{\tau\psi}^2
-
\frac{1}{\gamma_{0b}^2}\partial_{\psi}^2
+
k^2
\right]A
=
\frac{1}{\epsilon_0 c}\,J_z\,e^{-ikx},
\label{eq:1}
\end{equation}
where \(\gamma_{0b}=(1-v_{0b}^2/c^2)^{-1/2}\) is the relativistic Lorentz factor of the beam electrons. Linearising the relativistic equations of motion and the continuity equation for both the background-plasma and beam electrons, the perturbed quantities are obtained as

\begin{subequations}\label{eq:2}
\begin{align}
    & v_{1pz} = \frac{e}{mc}\,A\,e^{ikx}, \\
    &\partial_{\tau}\left[\gamma_{0b} \vec{v}_{1b} + \frac{\gamma_{0b}^3 v_{0b}^2 v_{1bz}}{c^2}\hat{z} \right] = -\frac{ikAv_{0b}e}{mc}\,e^{ikx}\hat{x} \nonumber \\
    &\qquad\qquad\qquad\quad\qquad\qquad\qquad+ \frac{e}{mc}\left[\partial_{\tau} + v_{0b} \partial_{\psi}\right]\vec{A}, \\
    &\partial_{\tau}^2 n_{1b} = \frac{n_{0b}e F(\psi)}{mc^2 \gamma_{0b}} \left[ -k^2 v_{0b} c + c \gamma_{0b}^{-2} \partial_{\psi}(\partial_{\tau} + v_{0b}\partial_{\psi}) \right. \nonumber \\
    &\qquad\qquad\qquad\quad\qquad\left. + G(\psi)(\partial_{\tau} + v_{0b}\partial_{\psi})\right] A\,e^{ikx},
\end{align}
\end{subequations}
    
where $G(\psi) = c\,\partial_\psi F(\psi)/[\gamma_{0b}^2 F(\psi)]$. Taking the second-order time derivative of Eq.~(\ref{eq:1}), substituting the perturbed quantities of Eq.~(\ref{eq:2}), and discarding only the highest-order derivative terms $(\partial_{\tau}^{4},\,\partial_{\tau\psi}^{4},\,\partial_{\psi}^{4})$ yields the extended evolution equation

\begin{widetext}

\begin{equation}\label{eq:3}
    \left[ \partial_{\tau}^2 + QF(\psi)\,\partial_{\tau\psi}^2 + P\,\partial_{\psi}^2 - \frac{Q}{2}\partial_{\psi}F(\psi)\,\partial_{\tau} - P\,\partial_{\psi}F(\psi)\,\partial_{\psi} - \Gamma_{0}^2 F(\psi) \right]A = 0,
\end{equation}
    
\end{widetext}
\begin{figure*}
    \centering
    \includegraphics[width=1\linewidth]{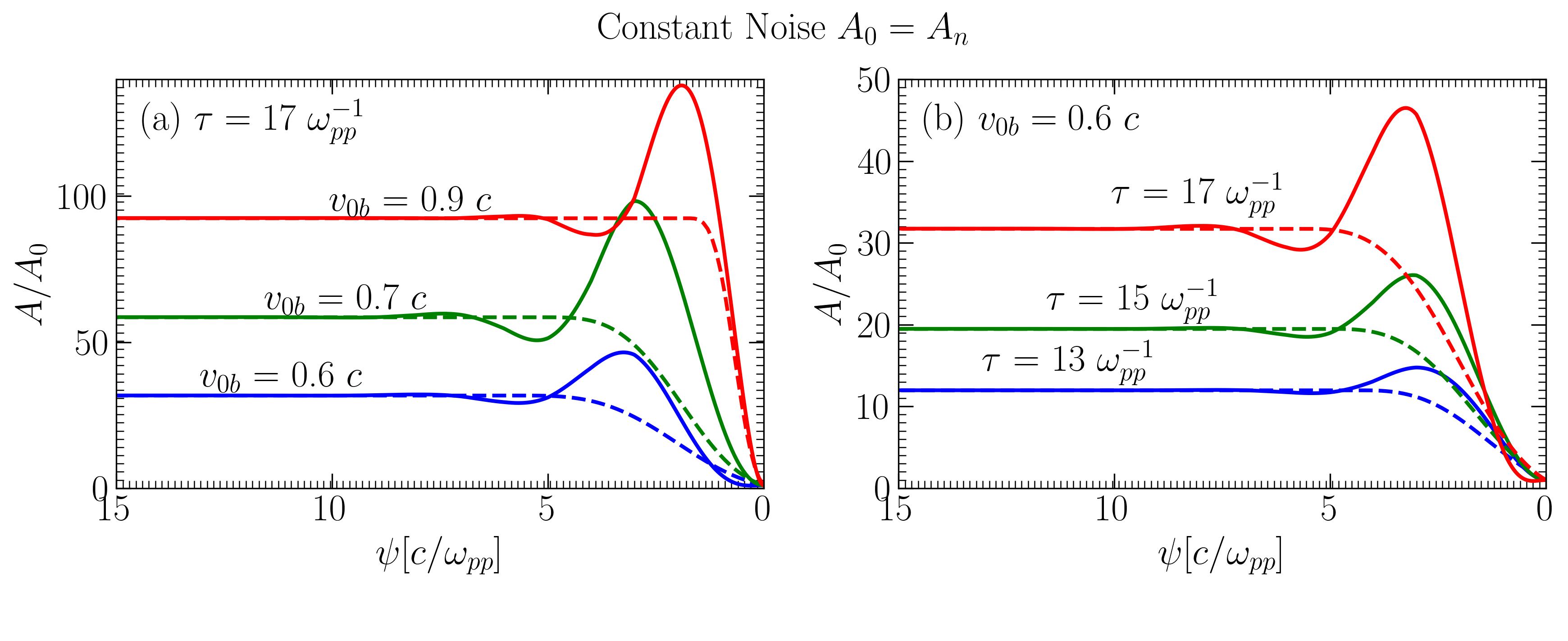}
    \caption{Comparison between the numerical solution of the present model (solid lines) and the analytical solution of Pathak \textit{et al.}~\cite{pathak2015spatial} (dashed lines) for constant initial noise $A_0 = A_n$. The present formulation includes the additional $P\partial_{\psi}^2$ term, which introduces enhanced spatial modulation in the downstream region not captured by the $P=0$ model. Panel~(a) shows profiles at fixed time $\tau = 17\,\omega_{pp}^{-1}$ for three beam velocities; panel~(b) shows temporal evolution at fixed $v_{0b} = 0.6c$.}
    \label{fig:1}
\end{figure*}

where $Q = 2\omega_{pb}v_{0b}/\gamma_{0b}^3 D$ and $P = Qv_{0b}/2$ are the spatial-propagation coefficients and $\Gamma_0 = kv_{0b}\omega_{pb}/\sqrt{\gamma_{0b}D}$ is the purely temporal growth rate in the absence of spatial coupling. Here $\omega_{pb} = (n_{0b}e^2/m\varepsilon_0)^{1/2}$ is the beam plasma frequency and $D$ is the dispersion denominator defined in Ref.~\cite{pathak2015spatial}. For a sharp beam front, $F(\psi) = 1$, the gradient terms vanish and Eq.~(\ref{eq:3}) reduces to

\begin{equation}\label{eq:4}
    \left[ \partial_{\tau}^2 + Q\,\partial_{\tau\psi}^2 + P\,\partial_{\psi}^2 - \Gamma_{0}^2 \right]A = 0.
\end{equation}

For $P = 0$, Eq.~\eqref{eq:4} further reduces to the leading-order formulation of Pathak \textit{et al.}~\cite{pathak2015spatial},

\begin{equation}\label{eq:4b}
    \left[ \partial_{\tau}^2 + Q\,\partial_{\tau\psi}^2 - \Gamma_{0}^2 \right]A = 0.
\end{equation}
To obtain the reference solution against which the extended model is later assessed, we apply a double Laplace transform in the co-moving coordinates $(\tau, \psi)$, defined as
\begin{equation}
\tilde{A}(\alpha, \beta) = L(A(\alpha, \beta))=\int_0^{\infty} d\tau \int_0^{\infty} d\psi\; 
A(\tau, \psi)\, e^{-i\alpha\tau - i\beta\psi},
\label{eq:5}
\end{equation}
where $\alpha$ and $\beta$ are the spectral variables conjugate to $\tau$ and $\psi$, respectively. Applying this transform to the derivatives gives

\begin{subequations}
    \begin{align}
    & L(\partial_{\tau}^2 A) = - \partial_{\tau} \tilde{A}(0, \beta) - i \alpha \tilde{A}(0, \beta) - \alpha^2 \tilde{A}(\alpha, \beta) \\
    & L(\partial_{\tau\psi}^2A) = -\partial_\psi \tilde{A}(0, \beta) - i \alpha \tilde{A}(\alpha, 0) - \alpha \beta \tilde{A}(\alpha, \beta) \\ 
    & L(\partial_{\psi}^2 A) = -\partial_{\psi} \tilde{A}(\alpha, 0) -i\beta \tilde{A}(\alpha, 0) - \beta^2 \tilde{A}(\alpha, \beta) 
    \end{align}
    \label{eq:ld}
\end{subequations}
where $A(0, \beta) = L(A(0, \psi))$ and $A(\alpha, 0) = L(A(\tau, 0))$. Substituting Eq.~\eqref{eq:ld} into
Eq.~\eqref{eq:4} and invoking the standard initial-value properties of the Laplace transform reduces the partial differential equation to the algebraic relation
\begin{widetext}
    
\begin{equation}
\tilde{A}(\alpha,\beta) = -\frac{\partial_\tau \tilde{A}(0,\beta) 
+ Q \partial_\psi \tilde{A}(0,\beta) 
+ P \partial_\psi \tilde{A}(\alpha, 0)
+ i\alpha \tilde{A}(0,\beta) 
+ i(Q\alpha + P\beta)\tilde{A}(\alpha,0)}
{\alpha^2 + Q\alpha\beta + \Gamma_0^2 + P \beta^2}.
\label{eq:6}
\end{equation}
\end{widetext}
Equation~(\ref{eq:6}) expresses the transformed amplitude entirely in terms of the data prescribed on the two boundaries of the co-moving domain: the initial field at $\tau = 0$, carried by $\tilde{A}(0,\beta)$ and $\partial_\tau\tilde{A}(0,\beta)$, and the inflow field at the beam front $\psi = 0$, carried by $\tilde{A}(\alpha,0)$ and $\partial_\psi\tilde{A}(\alpha,0)$. Each physical situation therefore corresponds to a particular choice of these boundary terms, and the spatio-temporal evolution of the field is recovered by substituting that choice into Eq.~(\ref{eq:6}) and inverting the transform. In the following sections we evaluate this solution for a sequence of inflow-noise profiles, beginning with the simplest case of a steady noise entering at the beam front and turning subsequently to time-dependent profiles in Sec.~\ref{sec:linosc}.

\section{Constant Noise}
We first consider a steady inflow, for which the field is constant on both boundaries. Imposing the constant initial amplitudes $A(0,\psi) = A_n$ and $A(\tau,0) = A_n$, with all derivatives set to zero, Eq.~(\ref{eq:6}) simplifies to
\begin{equation}
\tilde{A}(\alpha,\beta) = -\frac{1}{\alpha\beta}\frac{\alpha^2 + Q\alpha\beta + P\beta^2}{\alpha^2 + Q\alpha\beta + P\beta^2 + \Gamma_0^2}.
\label{eq:8}
\end{equation}

\begin{figure*}
    \centering
    \includegraphics[width=1\linewidth]{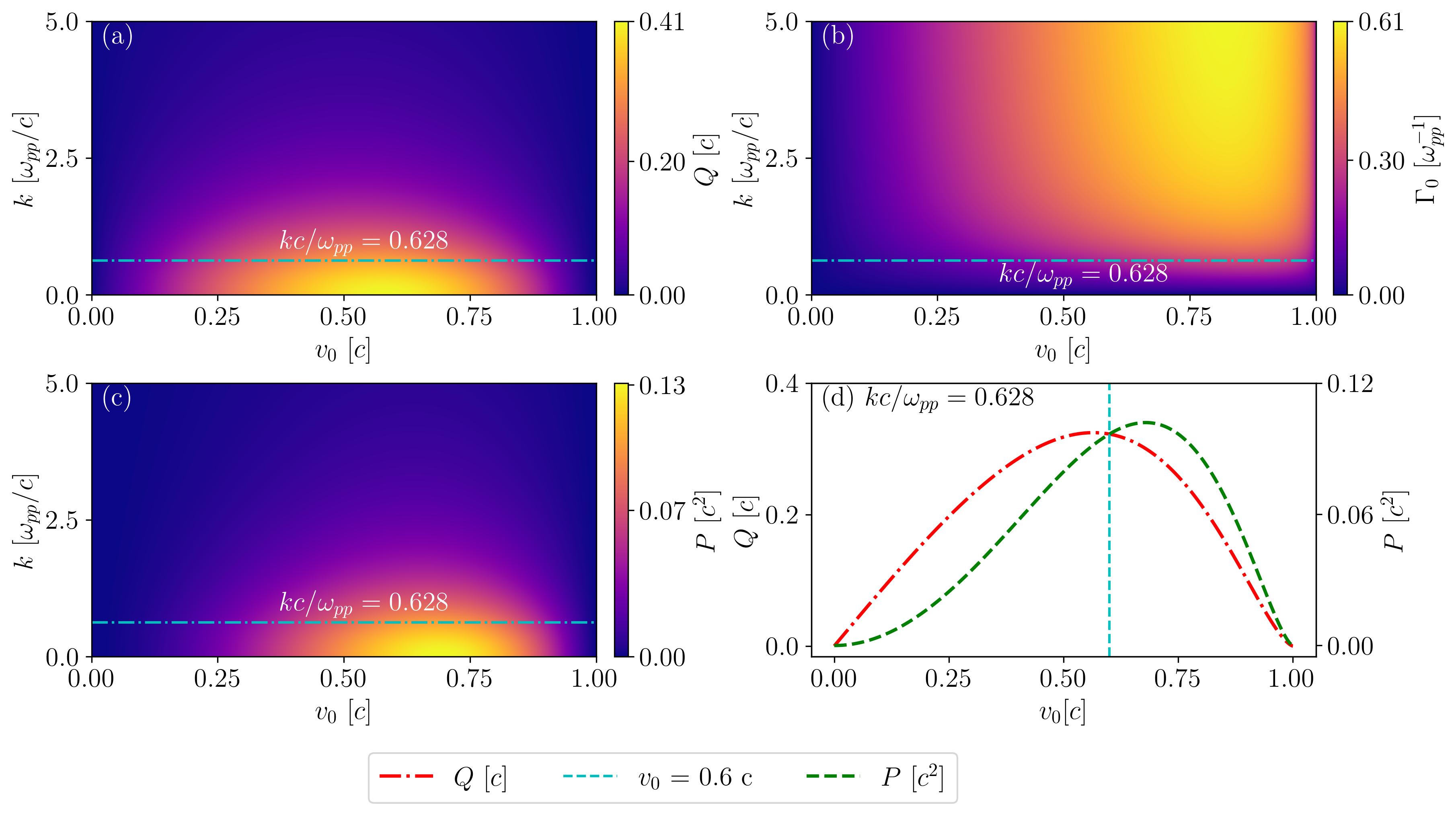}
    \caption{Spatial and temporal parameters of current filamentation instability as functions of wave vector $k$ and beam velocity $v_0$. Panels~(a) and~(c) show the spatial propagation coefficients $Q$ and $P$ in the $(k,v_0)$ plane, respectively. Panel~(b) shows the purely temporal growth rate $\Gamma_0(k,v_0)$. Panel~(d) shows profiles of $Q$ and $P$ versus $v_0$ at the working wave number $kc/\omega_{pp} = 0.628$ (cyan dashed line in panels a--c). Increasing $k$ suppresses spatial coupling while enhancing $\Gamma_0$, motivating the choice $kc/\omega_{pp} = 0.628$ for studying spatio-temporal effects.}
    \label{pqevo}
\end{figure*}

The physical field is recovered through the inverse transformation
\begin{equation}
A(\tau,\psi) = \frac{1}{4\pi^2}
\int_{-\infty-i\sigma_\alpha}^{\infty-i\sigma_\alpha} d\alpha
\int_{-\infty-i\sigma_\beta}^{\infty-i\sigma_\beta} d\beta \,
\tilde{A}(\alpha,\beta)\,e^{i\alpha\tau+i\beta\psi},
\label{eq:9}
\end{equation}
where $\sigma_\alpha$ and $\sigma_\beta$ are positive constants placing the integration contours to the right of all singularities. Since the inverse Laplace transform of Eq.~(\ref{eq:9}) cannot be obtained in closed analytical form, we set $P = 0$, following Pathak \textit{et al.}~\cite{pathak2015spatial}, in which case the solution is~\cite{pathak2015spatial,abramowitz1965handbook}
\begin{equation}
\begin{aligned}
&A(\tau,\psi) = S_n(\tau,\psi)
= A_n\Bigg[
\bigl(H(\tau) - H(\tau-\psi/Q)\bigr)\cosh(\Gamma_0\tau)\\
&+ H(\tau-\psi/Q)
\sum_{j=0}^{\infty}
\left(\frac{\psi/Q}{\tau-\psi/Q}\right)^j
I_{2j}\!\left(2\Gamma_0\sqrt{\frac{\psi}{Q}\!\left(\tau - \frac{\psi}{Q}\right)}\right)
\Bigg],
\end{aligned}
\label{eq:10}
\end{equation}
where 
\[
H(x)=
\begin{cases}
0, & x < 0 \\
1, & x \ge 0
\end{cases}
\] is the Heaviside step function and $I_{2j}$ is the modified Bessel function of the first kind of order $2j$~\cite{abramowitz1965handbook}. Equation~(\ref{eq:10}) serves as the $P = 0$ reference solution; the full numerical solution of Eq.~(\ref{eq:4}) with $P = Qv_{0b}/2$ is compared against it in Fig.~(\ref{fig:1}).

Figure~(\ref{fig:1}) compares the analytical solution, Eq.~(\ref{eq:10}), with the numerical solution of the present model for different beam velocities and times. The numerical results reproduce the overall growth behaviour of the analytical solution, confirming the consistency of the numerical implementation. They also display clear modifications of the spatial growth profile, most evident in the downstream region, where enhanced spatial modulation and localised amplitude variation appear. These features originate directly from the $P\,\partial_\psi^2 A$ term in Eq.~(\ref{eq:4}) and are absent from the $P = 0$ analytical expression.

Figure~(\ref{pqevo}) shows how the spatial coefficients $Q$ and $P$ and the growth rate $\Gamma_0$ vary with the wave vector $k$ and the beam velocity $v_0$. Increasing $k$ suppresses both spatial coefficients while raising $\Gamma_0$, so the spatio-temporal regime is most prominent at small $k$. The working wave number $kc/\omega_{pp} = 0.628$ is chosen because the spatial--temporal coupling is significant there; combined with $v_{0b} = 0.6c$ it gives $P/Q = 0.3$, confirming that the second-order correction is non-negligible.

Having established the analytical reference and the structure of the numerical solution, we now test the modified model against self-consistent kinetic simulations. Two-dimensional particle-in-cell (PIC) simulations were performed with the OSIRIS framework~\cite{fonseca2002} under conditions favourable to the current filamentation instability. A charge-neutral electron--proton beam with relativistic velocity $v_{0b} = 0.75\,c$ propagates through an initially cold, uniform background plasma. All species are initialised with zero thermal velocity, so that instability growth is governed entirely by the electromagnetic beam--plasma drive, with no thermal correction to the growth rate~\cite{califano1998,yang1994}.

\begin{figure*}
        \centering
        \includegraphics[width=1\linewidth]{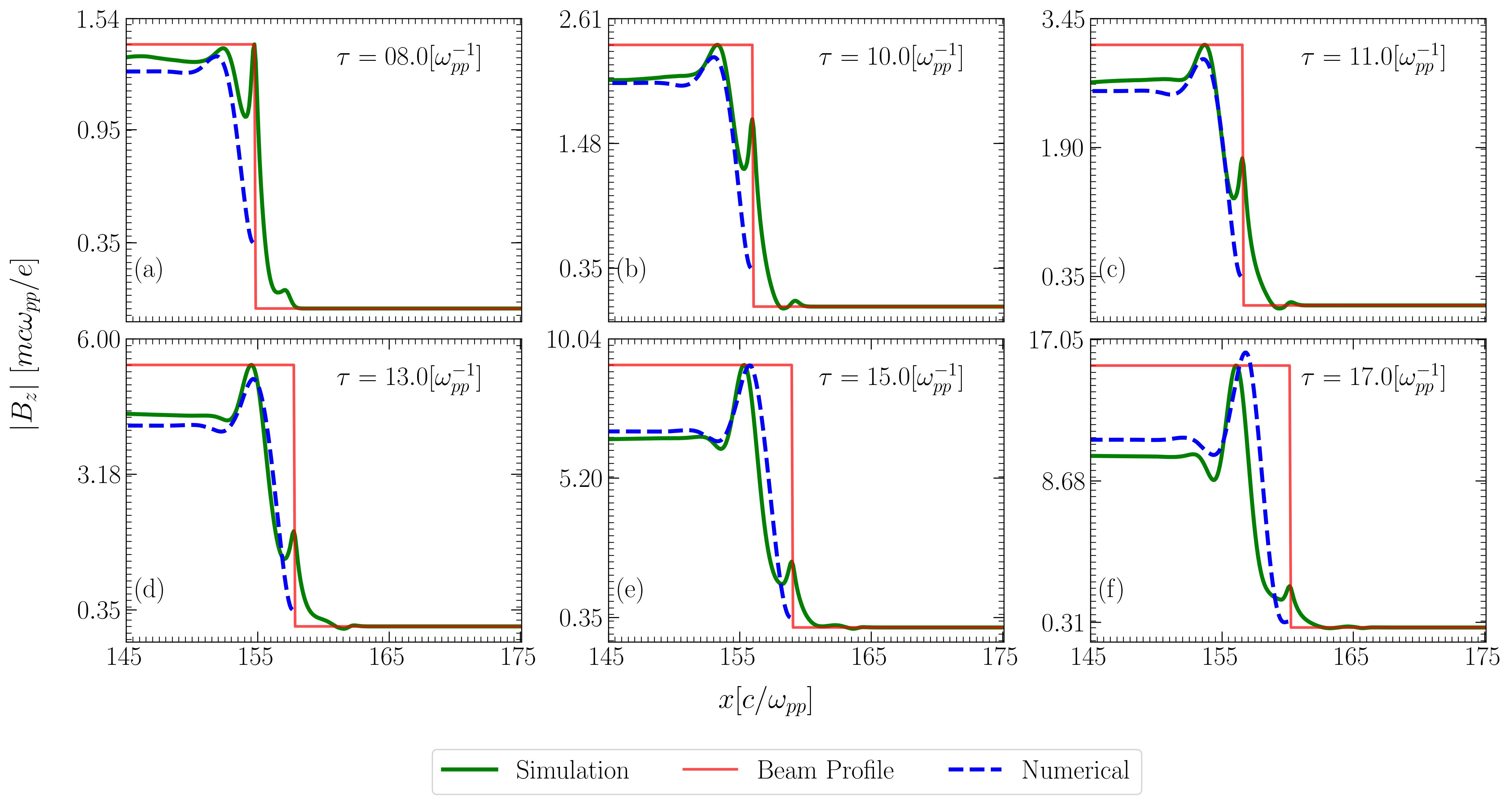}
        \caption{Temporal sequence of the magnetic-field amplitude $|B_z|(x)$ along the beam propagation axis for $v_{0b} = 0.6c$. Green solid: 2D PIC simulation; blue dashed: numerical solution of the modified model Eq.~(\ref{eq:4}); red solid: beam density profile. Panels~(a)--(f) correspond to $\tau = 8, 10, 11, 13, 15, 17\,\omega_{pp}^{-1}$. The magnetic peak migrates downstream with the instability saturation front and grows exponentially; the modified model tracks both its position and amplitude more accurately than the leading-order description.}
        \label{fig:3}
\end{figure*}
\begin{figure*}
    \centering
    \includegraphics[width=1\linewidth]{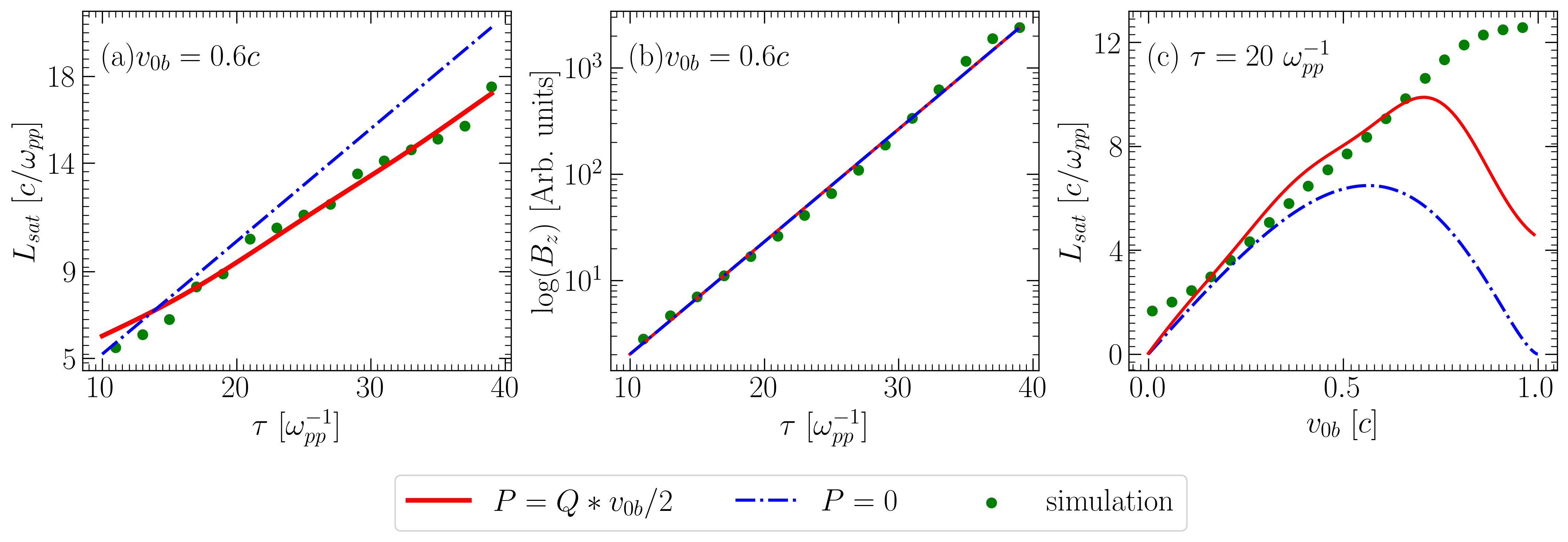}
    \caption{Comparison between the modified theory and 2D PIC simulations. (a)~Saturation length $L_{\mathrm{sat}}$ versus time at $v_{0b} = 0.6c$: red solid, modified prediction $(P+Q)\tau$; blue dash-dot, leading-order prediction $Q\tau$ ($P=0$); green dots, simulation. (b)~Logarithm of peak $|B_z|$ versus time confirming exponential temporal growth at rate $\Gamma_0$, unchanged by the $P$ correction. (c)~$L_{\mathrm{sat}}$ versus beam velocity at $\tau = 20\,\omega_{pp}^{-1}$; deviations above $v_{0b} \approx 0.6c$ indicate the onset of oblique-mode and nonlinear-filament dynamics.}
    \label{fig:4}
\end{figure*}
The simulation box measures $250\times100\,(c/\omega_{pp})^2$ and is discretised on a $2500\times1000$ grid, giving a uniform cell size $\Delta x = \Delta y = 0.1\,c/\omega_{pp}$ with $3\times3$ macro-particles per cell per species. A sharp beam front is imposed through a Heaviside density profile, which maintains a well-defined leading edge whose interaction with the unperturbed background plasma controls the onset of CFI. To select a single dominant transverse filamentation mode, a weak seed magnetic field $B_y(\tau=0) = B_0\cos(k_0 x)$ with $B_0 = 5\times10^{-5}\,mc\omega_{pp}/e$ is applied; this amplitude is small enough to leave the beam--plasma equilibrium unaffected while suppressing broadband numerical noise.

Figure~(\ref{fig:3}) shows the evolution of the magnetic-field component $B_z$ along the beam-propagation direction at successive times for $v_{0b} = 0.6c$. The red solid line marks the beam density profile, the green solid curve the PIC result, and the blue dashed curve the numerical solution of the modified model. At early times the magnetic-field growth is confined to a narrow region immediately behind the beam head, where the transverse current separation first develops. The peak of $|B_z|$ lies slightly behind the beam front because the background electrons require a finite electromagnetic response time before the return-current imbalance is fully established. This spatial confinement is the defining signature of the spatial-growth regime, in which the instability has not yet propagated far into the interaction region.

As the beam propagates further, the magnetic peak grows and shifts downstream, tracing the motion of the instability saturation front. The numerical solution of the modified model captures both the displacement of this peak and the gradual expansion of the active region more accurately than the $P = 0$ prediction, showing that the second-order spatial correction improves the description of beam-front evolution. A secondary magnetic structure just behind the main peak, visible at late times, is likewise reproduced qualitatively by the modified model through the additional spatial modulation carried by the $P\,\partial_\psi^2 A$ term.

The quantitative measure of this improvement is the saturation length $L_{\mathrm{sat}}$, defined as the distance from the beam front to the point at which spatial growth saturates and temporal amplification takes over. Figure~(\ref{fig:4}a) shows that $L_{\mathrm{sat}}$ grows approximately linearly with time and that the simulation data closely follow the modified prediction $(P+Q)\tau$. The slope $dL_{\mathrm{sat}}/d\tau \approx P+Q$ confirms that the higher-order spatial term sets the saturation boundary. Figure~(\ref{fig:4}b) shows the logarithm of the peak magnetic field $|B_z|$ measured beyond the saturation length as a function of time; the linear trend confirms exponential growth at the rate $\Gamma_0$ once the saturation front is established. This slope is identical for $P = 0$ and $P \neq 0$, verifying that the included term modifies the spatial transport without altering the temporal evolution.

Figure~(\ref{fig:4}c) shows the dependence of $L_{\mathrm{sat}}$ on beam velocity at the fixed time $\tau = 20\,\omega_{pp}^{-1}$. For $v_{0b} \leq 0.6c$ the simulation data agree well with the modified prediction. At higher Lorentz factors the simulated $L_{\mathrm{sat}}$ approaches a nearly constant value instead of decaying as the theory predicts. This deviation may arise because the sharp beam front generates strong local gradients near the beam edge that excite oblique modes with a longitudinal component not contained in the single-$k$ description \cite{bret2004,spitkovsky2008a}, or because at late times the plasma electrons begin to respond ahead of the formal beam boundary, so that the effective interaction region extends beyond $\psi = 0$ \cite{chang2008}. Both effects are captured in the simulation but lie outside the present model.

\begin{figure*}
    \centering
    \includegraphics[width=1\linewidth]{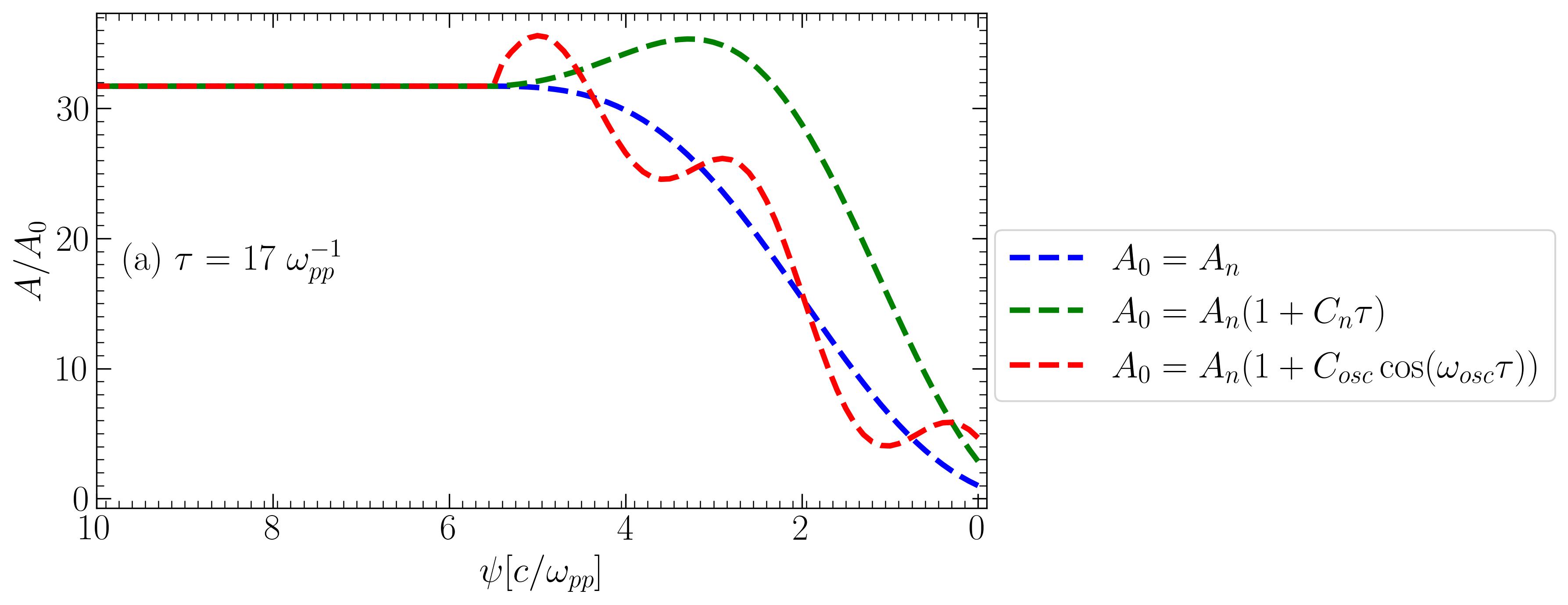}
    \caption{Analytical solutions for the three initial noise profiles at $\tau = 17\,\omega_{pp}^{-1}$, $v_{0b} = 0.6c$. Blue dashed: constant noise ($A_0 = A_n$); green dashed: linear noise ($A_0 = A_n(1+C_n\tau)$); red dashed: oscillatory noise ($A_0 = A_n(1+C_{\mathrm{osc}}\cos\omega_{\mathrm{osc}}\tau)$). All three solutions share the same saturation boundary at $\psi \approx L_{\mathrm{sat}}$; the oscillatory profile exhibits the richest spatial structure within the active region.}
    \label{fig:solution}
\end{figure*}

\section{Effect of Time-Dependent Noise}
\label{sec:linosc}

The constant-noise case treated above assumes that the perturbation entering at the beam front is steady. In realistic beam--plasma systems, however, the inflowing fluctuations may grow as the instability develops or vary periodically when the drive itself is modulated. To assess how the initial noise character imprints on the spatial field structure, we now solve the same evolution equation, Eq.~(\ref{eq:4}), for two further boundary conditions: a linearly growing noise and an oscillatory noise. The solution procedure is identical to that of Sec.~\ref{sec:const}; only the boundary term at $\psi = 0$ changes, so each result reduces to the constant-noise solution $S_n(\tau,\psi)$ plus an additional contribution set by the noise profile.

\textit{Case I: Linearly growing noise.}
The boundary value is taken as $A(\tau,0) = A_n(1 + C_{\mathrm{lin}}\tau)$ with $A(0,\psi) = A_n$, so that $dA(\tau,0)/d\tau = A_n C_{\mathrm{lin}}$. The Laplace-transformed equation becomes
\begin{equation}
\label{eq:11}
    A(\alpha, \beta) = -\frac{1}{\alpha\beta}\frac{\alpha^2 + Q\alpha\beta}{\alpha^2 + Q\alpha\beta + \Gamma_0^2} + \frac{iQ}{\alpha}\frac{A_n C_{\mathrm{lin}}}{\alpha^2 + Q\alpha\beta + \Gamma_0^2},
\end{equation}
and the inverse transform yields
\begin{widetext}
\begin{equation}
  \label{eq:12}
    A(\tau, \psi) = S_n(\tau,\psi) + H\!\left(\tau - \frac{\psi}{Q}\right)A_n C_{\mathrm{lin}}\left(\tau - \frac{\psi}{Q}\right) I_0\!\left[2\Gamma_0\sqrt{\frac{\psi}{Q}\!\left(\tau - \frac{\psi}{Q}\right)}\right].
\end{equation}
\end{widetext}
The additional term grows in proportion to the retarded time $\tau - \psi/Q$, weighted by a Bessel factor, and progressively amplifies the spatial field structure as the noise coefficient $C_{\mathrm{lin}}$ increases.

\textit{Case II: Oscillatory noise.}
For $A(\tau,0) = A_n[1 + C_{\mathrm{osc}}\cos(\omega_{\mathrm{osc}}\tau)]$, the Laplace transform acquires poles at $\alpha = \pm\omega_{\mathrm{osc}}$:
\begin{equation}
\label{eq:13}
    A(\alpha, \beta) = -\frac{1}{\alpha\beta}\frac{\alpha^2 + Q\alpha\beta}{\alpha^2 + Q\alpha\beta + \Gamma_0^2} - \frac{\alpha^2}{(\alpha^2 - \omega_{\mathrm{osc}}^2)}\frac{QA_n C_{\mathrm{osc}}}{\alpha^2 + Q\alpha\beta + \Gamma_0^2}.
\end{equation}
Applying the inverse laplace transformation, the solution becomes
\begin{widetext}
\begin{equation}
    \label{eq:14}
    A(\tau, \psi) = S_n(\tau,\psi) + H\!\left(\tau - \frac{\psi}{Q}\right)A_n C_{\mathrm{osc}}\cos\!\left[\omega_{\mathrm{osc}}\!\left(\tau - \frac{\psi}{Q}\right) - \frac{\Gamma_0^2}{Q}\frac{\psi}{\omega_{\mathrm{osc}}}\right].
\end{equation}
\end{widetext}
The cosine carries a spatial phase shift $-\Gamma_0^2\psi/(Q\omega_{\mathrm{osc}})$ that stretches the oscillation pattern along $\psi$ relative to its temporal period, a direct consequence of the growth-rate modulation.
\begin{figure*}
    \includegraphics[width=\linewidth]{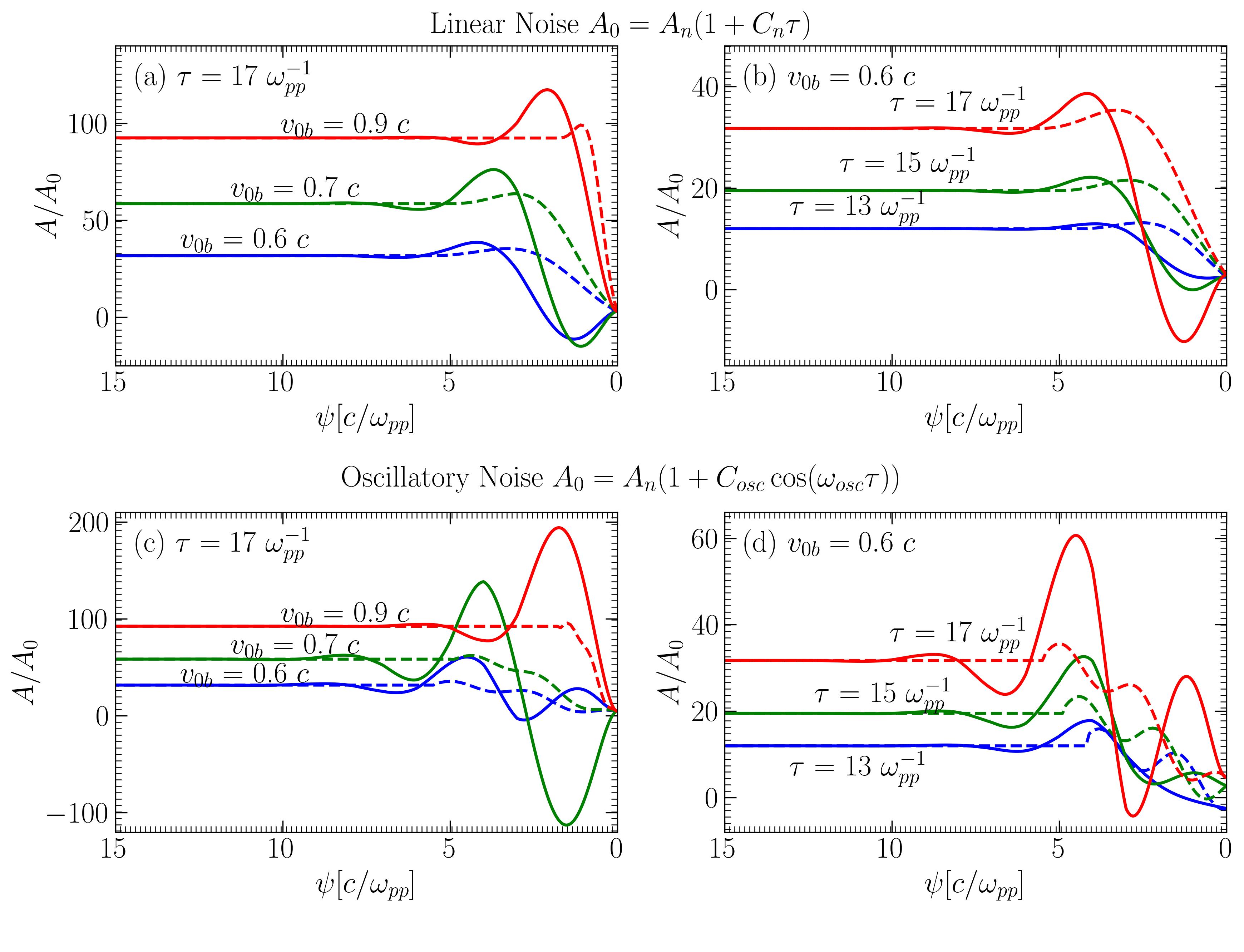}
    \caption{Comparison of field evolution for $P = 0$ (dashed lines) and $P = Qv_{0b}/2$ (solid lines) for linear noise with $C_{lin} = 0.11$ (panels a--b) and oscillatory noise with $C_{\mathrm{osc}} = 4$ and $\omega_{\mathrm{osc}} = 0.763$ (panels c--d). Left column: fixed $\tau = 17\,\omega_{pp}^{-1}$, varying $v_{0b}$. Right column: fixed $v_{0b} = 0.6c$, varying $\tau$. The higher-order term broadens the active spatial region and shifts local amplitude extrema, most prominently at high velocity and late time.}
    \label{numer}
\end{figure*}

Figure~(\ref{fig:solution}) shows the three analytical solutions at $\tau = 17\,\omega_{pp}^{-1}$ with $v_{0b} = 0.6c$. The field grows spatially up to the saturation length $L_{\mathrm{sat}} = Q\tau$ (the $P=0$ limit), beyond which the evolution becomes purely temporal; this saturation length is the same for all three profiles. As the perturbation amplitude of the initial noise increases, the spatial modulation of the field within the active region becomes more pronounced, but the saturation length $L_{\mathrm{sat}}$ is unchanged.

Figure~(\ref{numer}) compares the $P = 0$ (dashed) and $P = Qv_{0b}/2$ (solid) numerical solutions for linear and oscillatory noise over a range of beam velocities and times. Including the higher-order derivative $\partial_\psi^2$ broadens the active spatial region, shifts the positions of the local amplitude extrema, and, for the oscillatory case, modifies the spatial phase of the downstream oscillations. With the higher-order correction the effective saturation length is \[L_{\mathrm{sat}} \approx (P+Q)\tau .\] Using \(P = Qv_{0b}/2\), this becomes \[L_{\mathrm{sat}} \approx Q\left(1+\frac{v_{0b}}{2}\right)\tau =\frac{(2 + v_{0b})\omega_{pb}v_{0b}}{\gamma_{0b}^3 D}\tau,
\] so that the saturation region is enlarged by a factor \(v_{0b}/2\) relative to the leading-order value \(Q\tau\), corresponding to $30\%$ increase for $v_{0b}=0.6c$. Beyond $v_{0b} \approx 0.6c$ the saturation length decreases, because the $\gamma_{0b}^3$ weighting in the denominator of $Q$ rapidly compresses both spatial coefficients and the downstream field becomes dominated by purely temporal growth.

\section{Discussion and Conclusion}

The extended spatio-temporal model derived here includes one additional spatial degree of freedom---the $P\,\partial_\psi^2 A$ term---relative to the leading-order formulation of Pathak \textit{et al.}~\cite{pathak2015spatial}. This term represents the second-order contribution to spatial propagation of the instability front arising from the $v_{0b}^2$ correction to the beam electron dynamics, which is negligible at non-relativistic velocities but reaches 30\% of the leading-order coefficient at $v_{0b} = 0.6c$. The most direct consequence is that the saturation front advances at speed $P + Q = Q(1 + v_{0b}/2)$ rather than $Q$, enlarging the spatial interaction region and shifting the transition point between the spatial and temporal growth regimes.

This correction has practical importance for any application that depends on the physical size of the magnetized precursor region, including estimates of the mean free path for diffusive shock acceleration \cite{achterberg2007}, the spatial extent of the Weibel shock ramp \cite{spitkovsky2008a,chang2008}, and the threshold condition for filament merging and inverse magnetic-energy cascade \cite{silva2003,stockem2014}. The choice $v_{0b} = 0.6c$ is also physically motivated: subrelativistic bulk velocities in this range characterize the deceleration phase of certain GRB ejecta components and the downstream regions of mildly relativistic AGN jets, where CFI-driven magnetic fields are thought to mediate collisionless shock formation and non-thermal particle acceleration \cite{gehrels2002brightest,spitkovsky2008particle,medvedev1999,frederiksen2004,hededal2004}.

The three analytical solutions further show that the spatial structure of the field within the active region carries an imprint of the initial noise character. Oscillatory noise produces the richest downstream structure, with a spatial oscillation frequency set by $\Gamma_0^2/(Q\omega_{\mathrm{osc}})$. This dependence means that analytical estimates based on constant or linear noise assumptions may underestimate the spatial variance of the magnetic field when the actual inflow fluctuations are periodic, as can occur in laser-driven beam--plasma experiments where the laser envelope introduces a natural modulation frequency \cite{fiuza2020electron}.

The failure of the modified model at $v_{0b} \gtrsim 0.7c$ points to physics that lies beyond the present single-mode reduced description. Oblique modes, which combine two-stream and filamentation drives and become competitive at high Lorentz factors, redirect free energy away from the purely transverse channel \cite{bret2004,bret2010}. Additionally, nonlinear filament coalescence at late times $\tau \gtrsim 30\,\omega_{pp}^{-1}$ generates an inverse cascade in the magnetic spectrum that invalidates the single-$k$ assumption underlying the partial differential equation \cite{silva2003,stockem2014,kazimura1998}. Extending the model to cover these regimes would require either a multi-mode spectral formulation or a hybrid analytical--kinetic approach.

In summary, including the second-order spatial term $P\,\partial_\psi^2 A$ extends the saturation length by the factor $(1 + v_{0b}/2)$ and brings the analytical prediction into substantially improved quantitative agreement with the 2D PIC simulations for $v_{0b} \leq 0.6c$. The temporal growth rate $\Gamma_0$ is unaffected, confirming that the correction acts on spatial propagation rather than local amplification. Oscillatory initial noise produces the largest spatial modulation within the active region, with a characteristic frequency controlled by $\Gamma_0^2/(Q\omega_{\mathrm{osc}})$. Deviations above $v_{0b} \approx 0.6c$ indicate that oblique modes and nonlinear filament dynamics become important and cannot be absorbed into a single-mode spatial correction. The modified model provides a quantitatively improved and computationally inexpensive description of CFI in the spatio-temporal regime, with direct relevance to both laboratory beam--plasma experiments and astrophysical contexts where the magnetic-field scale length governs shock structure and particle acceleration \cite{medvedev1999,nishikawa2009,fiuza2020electron,treumann2009}.

\begin{acknowledgments}
The authors would like to acknowledge the support from VIT SEED grant funding SG20220070 and partial support from the Department of Science and Technology, Government of India, under grant number SR/FST/MS-II/2023/139(C)-VIT Vellore.
\end{acknowledgments}

\bibliographystyle{apsrev4-2}
\bibliography{referance_new}

@article{bret2010multidimensional,
  title={Multidimensional electron beam-plasma instabilities in the relativistic regime},
  author={Bret, Antoine and Gremillet, Laurent and Dieckmann, Mark Eric},
  journal={Physics of Plasmas},
  volume={17},
  number={12},
  pages={120501},
  year={2010},
  publisher={AIP Publishing}
}

@article{pathak2015spatial,
  title={Spatial-temporal evolution of the current filamentation instability},
  author={Pathak, Vishwa Bandhu and Grismayer, Thomas and Stockem, Anne and Fonseca, R. A. and Silva, L. O.},
  journal={New Journal of Physics},
  volume={17},
  number={4},
  pages={043049},
  year={2015},
  publisher={IOP Publishing}
}

@book{abramowitz1965handbook,
  title={Handbook of Mathematical Functions: With Formulas, Graphs, and Mathematical Tables},
  author={Abramowitz, Milton and Stegun, Irene A.},
  volume={55},
  year={1965},
  publisher={Courier Corporation}
}

@article{gehrels2002brightest,
  title={The brightest explosions in the universe},
  author={Gehrels, Neil and Piro, Luigi and Leonard, Peter J. T.},
  journal={Scientific American},
  volume={287},
  number={6},
  pages={84--91},
  year={2002},
  publisher={JSTOR}
}

@article{spitkovsky2008particle,
  title={Particle acceleration in relativistic collisionless shocks: Fermi process at last?},
  author={Spitkovsky, Anatoly},
  journal={The Astrophysical Journal},
  volume={682},
  number={1},
  pages={L5},
  year={2008},
  publisher={IOP Publishing}
}

@article{fiuza2020electron,
  title={Electron acceleration in laboratory-produced turbulent collisionless shocks},
  author={Fiuza, F. and Swadling, G. F. and Grassi, A. and Rinderknecht, H. G. and Higginson, D. P. and Ryutov, D. D. and Bruulsema, C. and Drake, R. P. and Funk, S. and Glenzer, S. and others},
  journal={Nature Physics},
  volume={16},
  number={9},
  pages={916--920},
  year={2020},
  publisher={Nature Publishing Group UK London}
}

@article{weibel1959,
  title={Spontaneously Growing Transverse Waves in a Plasma Due to an Anisotropic Velocity Distribution},
  author={Weibel, Erich S.},
  journal={Physical Review Letters},
  volume={2},
  number={3},
  pages={83--84},
  year={1959},
  publisher={American Physical Society}
}

@article{fried1959,
  author={Fried, Burton D.},
  title={Mechanism for Instability of Transverse Plasma Waves},
  journal={The Physics of Fluids},
  volume={2},
  number={3},
  pages={337--337},
  year={1959}
}

@article{rosenbluth1957,
  author={Rosenbluth, M. N. and Longmire, C. L.},
  title={Stability of plasmas confined by magnetic fields},
  journal={Annals of Physics},
  volume={1},
  number={2},
  pages={120--140},
  year={1957}
}

@article{bret2009,
  year={2009},
  publisher={The American Astronomical Society},
  volume={699},
  number={2},
  pages={990},
  author={Bret, A.},
  title={{Weibel, Two-Stream, Filamentation, Oblique, Bell, Buneman\ldots\ Which One Grows Faster?}},
  journal={The Astrophysical Journal}
}

@article{bret2010,
  title={Exact relativistic kinetic theory of the full unstable spectrum of an electron-beam--plasma system with {M}axwell-{J}{\"u}ttner distribution functions},
  author={Bret, A. and Gremillet, L. and B{\'e}nisti, D.},
  journal={Physical Review E},
  volume={81},
  pages={036402},
  year={2010},
  publisher={American Physical Society}
}

@article{yoon1987,
  title={Exact analytical model of the classical {W}eibel instability in a relativistic anisotropic plasma},
  author={Yoon, Peter H. and Davidson, Ronald C.},
  journal={Physical Review A},
  volume={35},
  pages={2718--2721},
  year={1987},
  publisher={American Physical Society}
}

@article{silva2003,
  year={2003},
  month={9},
  volume={596},
  number={1},
  pages={L121},
  author={Silva, L. O. and Fonseca, R. A. and Tonge, J. W. and Dawson, J. M. and Mori, W. B. and Medvedev, M. V.},
  title={Interpenetrating Plasma Shells: Near-Equipartition Magnetic Field Generation and Nonthermal Particle Acceleration},
  journal={The Astrophysical Journal}
}

@book{melrose1986,
  author={Melrose, D. B.},
  title={Instabilities in Space and Laboratory Plasmas},
  publisher={Cambridge University Press},
  address={Cambridge},
  year={1986},
  isbn={9780521305419}
}

@article{medvedev2005,
  year={2004},
  volume={618},
  number={2},
  pages={L75},
  author={Medvedev, Mikhail V. and Fiore, Massimiliano and Fonseca, Ricardo A. and Silva, Luis O. and Mori, Warren B.},
  title={Long-Time Evolution of Magnetic Fields in Relativistic Gamma-Ray Burst Shocks},
  journal={The Astrophysical Journal}
}

@article{nishikawa2009,
  year={2009},
  month={5},
  publisher={The American Astronomical Society},
  volume={698},
  number={1},
  pages={L10},
  author={Nishikawa, K.-I. and Niemiec, J. and Hardee, P. E. and Medvedev, M. and Sol, H. and Mizuno, Y. and Zhang, B. and Pohl, M. and Oka, M. and Hartmann, D. H.},
  title={{W}eibel Instability and Associated Strong Fields in a Fully Three-Dimensional Simulation of a Relativistic Shock},
  journal={The Astrophysical Journal}
}

@article{sironi2011,
  year={2010},
  publisher={The American Astronomical Society},
  volume={726},
  number={2},
  pages={75},
  author={Sironi, Lorenzo and Spitkovsky, Anatoly},
  title={Particle Acceleration in Relativistic Magnetized Collisionless Electron-Ion Shocks},
  journal={The Astrophysical Journal}
}

@article{medvedev1999,
  title={Generation of Magnetic Fields in the Relativistic Shock of Gamma-Ray Burst Sources},
  author={Medvedev, Mikhail V. and Loeb, Abraham},
  journal={The Astrophysical Journal},
  volume={526},
  number={2},
  pages={697--706},
  year={1999},
  publisher={IOP Publishing}
}

@article{fonseca2002,
  title={{OSIRIS}: A Three-Dimensional, Fully Relativistic Particle in Cell Code for Modeling Plasma Based Accelerators},
  author={Fonseca, R. A. and Silva, L. O. and Tsung, F. S. and Decyk, V. K. and Lu, W. and Ren, C. and Mori, W. B. and Deng, S. and Lee, S. and Katsouleas, T. and Adam, J. C.},
  journal={Lecture Notes in Computer Science},
  volume={2331},
  pages={342--351},
  year={2002},
  publisher={Springer}
}

@article{stockem2014,
  title={Relativistic {W}eibel instability},
  author={Stockem, A. and Dieckmann, M. E. and Schlickeiser, R.},
  journal={Plasma Physics and Controlled Fusion},
  volume={56},
  number={12},
  pages={125002},
  year={2014},
  publisher={IOP Publishing}
}

@article{davidson1972,
  title={Nonlinear development of electromagnetic instabilities in anisotropic plasmas},
  author={Davidson, Ronald C. and Hammer, David A. and Haber, I. and Wagner, C. E.},
  journal={The Physics of Fluids},
  volume={15},
  number={2},
  pages={317--333},
  year={1972},
  publisher={AIP Publishing}
}

@article{califano1998,
  title={Kinetic saturation of the {W}eibel instability in a collisionless plasma},
  author={Califano, F. and Prandi, R. and Pegoraro, F. and Bulanov, S. V.},
  journal={Physical Review E},
  volume={58},
  number={6},
  pages={7837--7845},
  year={1998},
  publisher={American Physical Society}
}

@article{yang1994,
  title={Kinetic evolution of the {W}eibel instability},
  author={Yang, T.-Y. B. and Gallant, Y. and Arons, J. and Langdon, A. B.},
  journal={Physics of Plasmas},
  volume={1},
  number={9},
  pages={3059--3077},
  year={1994},
  publisher={AIP Publishing}
}

@article{kazimura1998,
  title={Generation of a Small-Scale Quasi-Static Magnetic Field and Fast Particles during the Collision of Electron-Positron Plasma Clouds},
  author={Kazimura, Y. and Sakai, J. I. and Neubert, T. and Bulanov, S. V.},
  journal={The Astrophysical Journal},
  volume={498},
  number={2},
  pages={L183--L186},
  year={1998},
  publisher={IOP Publishing}
}

@article{chang2008,
  title={A Three-Zone Model for the Dynamical Evolution of Relativistic Collisionless Shocks},
  author={Chang, Philip and Spitkovsky, Anatoly and Arons, Jonathan},
  journal={The Astrophysical Journal},
  volume={674},
  number={1},
  pages={378--387},
  year={2008},
  publisher={IOP Publishing}
}

@article{achterberg2007,
  title={Particle acceleration by ultrarelativistic shocks: theory and simulations},
  author={Achterberg, A. and Wiersma, J. and Norman, C. A.},
  journal={Astronomy \& Astrophysics},
  volume={475},
  number={1},
  pages={1--18},
  year={2007},
  publisher={EDP Sciences}
}

@article{frederiksen2004,
  title={Magnetic Field Generation in Collisionless Shocks: Pattern Growth and Transport},
  author={Frederiksen, J. T. and Hededal, C. B. and Haugb{\o}lle, T. and Nordlund, {\AA}.},
  journal={The Astrophysical Journal},
  volume={608},
  number={1},
  pages={L13--L16},
  year={2004},
  publisher={IOP Publishing}
}

@article{hededal2004,
  title={Non-Fermi Power-Law Acceleration in Astrophysical Plasma Shocks},
  author={Hededal, C. B. and Haugb{\o}lle, T. and Frederiksen, J. T. and Nordlund, {\AA}.},
  journal={The Astrophysical Journal},
  volume={617},
  number={2},
  pages={L107--L110},
  year={2004},
  publisher={IOP Publishing}
}

@article{bret2004,
  title={Collective electromagnetic modes for beam-plasma interaction in the whole $k$ space},
  author={Bret, A. and Firpo, M.-C. and Deutsch, C.},
  journal={Physical Review E},
  volume={70},
  number={4},
  pages={046401},
  year={2004},
  publisher={American Physical Society}
}

@article{spitkovsky2008a,
  title={On the Structure of Relativistic Collisionless Shocks in Electron-Ion Plasmas},
  author={Spitkovsky, Anatoly},
  journal={The Astrophysical Journal},
  volume={673},
  number={1},
  pages={L39--L42},
  year={2008},
  publisher={IOP Publishing}
}

@article{treumann2009,
  title={Fundamentals of collisionless shocks for astrophysical application, 1. Non-relativistic shocks},
  author={Treumann, R. A.},
  journal={The Astronomy and Astrophysics Review},
  volume={17},
  number={4},
  pages={409--535},
  year={2009},
  publisher={Springer}
}

\end{document}